\documentclass[12pt]{amsart}
\usepackage{amsmath,amsthm,amssymb}
  \def\P#1#2{G_{#1}^{(#2)}}
\def\pmatrix{\left(\begin{matrix}}
\def\endpmatrix{\end{matrix}\right)}
\def\Sp{\operatorname{Sp}}

\def\Z{{\mathbb Z}}
\def\F{{\mathbb Z}_2}
\def\C{{\mathbb C}}

\def\de{\delta}

\def\t{\theta}

\def\T{\Theta}
\def\e{\varepsilon}

\def\a{\alpha}
\def\b{\beta}
\def\A{{\mathcal A}}
\def\M{{\mathcal M}}

\def\H{{\mathcal H}}
\def\tch#1#2{{\left[\begin{matrix}#1\\ #2\end{matrix}\right]}}
\def\tt#1#2{{\t\tch{#1}{#2}}}
\theoremstyle{plain}
\newtheorem{thm}{Theorem}
\newtheorem{lm}[thm]{Lemma}
\newtheorem{prop}[thm]{Proposition}
\newtheorem{cor}[thm]{Corollary}

\theoremstyle{definition}
\newtheorem{df}[thm]{Definition}
\newtheorem{rem}[thm]{Remark}

\begin{document}
\title{Superstring scattering amplitudes in higher genus}
\author{Samuel Grushevsky}
\thanks{Research is supported in part by National Science Foundation under the grant DMS-05-55867}
\address{Mathematics Department, Princeton University, Fine Hall, Washington Road, Princeton, NJ 08544, USA}

\begin{abstract}
In this paper we continue the program pioneered by D'Hoker and Phong, and recently advanced by Cacciatori, Dalla Piazza, and van Geemen, of finding the chiral superstring measure by constructing modular forms satisfying certain factorization constraints. We give new expressions for their proposed ans\"atze in genera 2 and 3, respectively, which admit a straightforward generalization. We then propose an ansatz in genus 4 and verify that it satisfies the factorization constraints and gives a vanishing cosmological constant. We further conjecture a possible formula for the superstring amplitudes in any genus, subject to the condition that certain modular forms admit holomorphic roots.
\end{abstract}
\maketitle

\section{Introduction}
The problem of finding an explicit expression for the string measure to an arbitrary loop (genus) order is one of the major open problems in string perturbation theory. For the bosonic string, chiral expressions in any genus in terms of theta functions and additional points on the worldsheet have been proposed by Manin in \cite{Manin}, Beilinson and Manin in \cite{BM}, and Verlinde and Verlinde in \cite{VV}. Non-chiral expressions in terms of the Weil-Petersson measure and Selberg zeta functions have also been proposed by D'Hoker and Phong in \cite{DHP86}. The problem is much more difficult for the superstring and the heterotic string.
Although the one-loop scattering amplitudes had been derived
by Green and Schwarz in \cite {GS} for the superstring and by
Gross, Harvey, Martinec, and Rohm in \cite{GHMR} for the heterotic string, the case of genus $g\geq 2$ remained inaccessible
until relatively recently. The difficulty in obtaining a formula for the chiral superstring measure in higher genus is the occurrence of odd supermoduli in any amplitude starting from genus $g\geq 2$.

\smallskip
A solution to this problem was proposed by D'Hoker and Phong, who introduced a gauge-fixing procedure respecting the local supersymmetry of the worldsheet.  Using this procedure, they managed to compute the genus 2 superstring measure from first principles
in a series of papers \cite{DHP1,DHPa,DHPb,DHPc} and to verify that the corresponding result $\theta[\Delta]^4\Xi_6[\Delta]$ times the bosonic measure produced vanishing cosmological constant, 2- and 3-point scattering amplitudes, and other expected physical properties \cite{DHPd,DHPe}. With the new insight from $\Xi_6[\Delta](\Omega)$, they started a modern program of identifying the higher genera superstring measure from factorization constraints and syzygy/asyzygy conditions \cite{DHP2, DHP3}. In \cite{DHP3} they also proposed an ansatz for the superstring measure in genus 3, including a $\theta[\Delta]^4$ factor, subject to the condition of certain linear combinations of modular forms having a square root. However, to date such a linear combination has not been found, and may not exist (see below).

\smallskip
In \cite{CDP} Cacciatori and Dalla Piazza used the  combinatorics of the action of the symplectic group on the set of theta characteristics in genus 2 to identify D'Hoker and Phong's
modular form $\Xi_6[\Delta](\Omega)$ from certain invariance properties. Recently Cacciatori, Dalla Piazza, and van Geemen \cite{CDPvG} proposed an ansatz for the chiral superstring measure in genus 3, by constructing an appropriate modular form $\Xi_8[\Delta]$, which has a $\theta[\Delta]^2$ rather than the $\theta[\Delta]^4$ factor, satisfying the factorization constraints on the locus of products of abelian varieties of lower genera. They also promise to determine in a forthcoming paper the dimension of the appropriate space of modular forms, and to show that their form is the unique one satisfying the factorization constraints (and thus there would be no solution in the form suggested in \cite{DHP3}). They also say that their constraints appear to have a solution in genus 4 as well.

\smallskip
In this paper we rewrite the ansatz of D'Hoker and Phong in genus 2, and of Cacciatori, Dalla Piazza, and van Geemen in genus 3 in terms of modular forms associated to isotropic spaces of theta characteristics, which have been studied since the times of Krazer \cite{K} and in particular used by Salvati Manni \cite{SM}. This allows us to propose a straightforward generalization of the chiral superstring measure to higher genera, which for genus 4 is an appropriate holomorphic modular form satisfying the necessary factorization constraints and producing vanishing cosmological constant. For higher genera we conjecture a possible ansatz, satisfying the factorization constraints, contingent on certain monomials in theta constants admitting holomorphic roots.

\smallskip
An expression for higher genus superstring amplitudes was also proposed by Matone and Volpato \cite{MV}. Their formulas depend on the choice of points on the worldsheet and do not seem to give an explicit modular form. It would be interesting to understand the relation of our work to theirs.

\smallskip
The structure of this work is as follows: in section 2 we fix notations and introduce basic notions of modular forms. In section 3 we review the orbits of the action of the symplectic group on sets of theta characteristics. In section 4 we reinterpret the modular form $G$ defined in \cite{CDPvG} in terms of syzygies. Though strictly speaking this computation is not needed to define our modular forms and construct an ansatz, this is our motivation for considering, in section 5, modular forms corresponding to vectors subspaces of the space of theta characteristics and reviewing what is known about them. In section 6 we prove the crucial theorem \ref{restrict} describing the restrictions of these modular forms to loci of products. In section 7 we obtain a new expression for the ans\"atze in genera 2 and 3 in terms of our modular forms, and also verify that in genus 4 there is a unique modular form that is a linear combination of ours that has correct factorization properties, thus giving an ansatz in genus 4. In section 8 we describe a possible generalization to arbitrary genus, proving in theorem \ref{Thmg} that it satisfies the factorization constraints (this also gives another proof of this for the ansatz in genus 4), and describe further tests and questions that could be used to study the validity and uniqueness of the ansatz.

\section{Notations and definitions}
\begin{df}
We denote by $\A_g$ the moduli space of complex principally polarized abelian varieties of dimension $g$. We denote by $\H_g$ the Siegel
upper half-space of symmetric complex matrices with positive-definite imaginary part, called period matrices. The right action of the symplectic group $\Sp(2g,\Z)$ on $\H_g$ is
given by
$$
  \pmatrix A&B\\  C&D\endpmatrix\circ\tau:= (C\tau+D)^{-1}(A\tau+B)
$$
where we think of elements of $\Sp(2g,\Z)$ as of consisting of four
$g\times g$ blocks, and they preserve the symplectic form given in
the block form as $\pmatrix 0& 1\\ -1& 0\endpmatrix$. We then have
$\A_g=\Sp(2g,\Z)\backslash\H_g$.
\end{df}
\begin{df}
Given a period matrix $\tau\in\H_g$ we denote the abelian variety
corresponding to $[\tau]\in\A_g$ by $A_\tau:=\C^g/(\Z^g+\tau\Z^g)$.
The theta function is a function of $\tau\in \H_g$ and $z\in\C^g$
given by
$$
  \theta(\tau,z):=\sum\limits_{n\in\Z^g}\exp(\pi i (n^t\tau n+2n^t
  z)).
$$
We denote by $\T_\tau$ the line bundle on $A_\tau$ of which the theta function is a section
\end{df}
\begin{df}
Given a point of order two on $A_\tau$, which can be uniquely
represented as $\frac{\tau\e+\de}{2}$ for $\e,\de\in \F^g$ (where $\F$ denotes the abelian group $\Z/2\Z=\lbrace 0,1\rbrace$ and we use the additive notations throughout the text), the associated
theta function with characteristic is
$$
  \tt\e\de(\tau,z):=\sum\limits_{n\in\Z^g}\exp(\pi i ((n+\e)^t\tau
  (n+\e)+ 2(n+\e)^t( z+\de)).
$$
As a function of $z$, $\tt\e\de$ is odd or even depending on whether
the scalar product $\e\cdot\de\in\F$ is equal to 1 or 0, respectively. Theta constants are restrictions of theta
functions to $z=0$, and all odd theta constants vanish identically in
$\tau$.
\end{df}
\begin{df}
A modular form of weight $k$ with respect to a subgroup
$\Gamma\subset \Sp(2g,\Z)$ is a function $f:\H_g\to\C^g$ such that
$$
  f(\gamma\circ\tau)=\det(C\tau+D)^kf(\tau)\quad
  \forall\gamma\in\Gamma,\forall\tau\in\H_g.
$$
We define the level subgroups of $\Sp(2g,\Z)$ as follows:
$$
  \Gamma_g(n):=\left\lbrace M=\pmatrix A&B\\ C&D\endpmatrix
  \in\Gamma_g\, |\, M\equiv\pmatrix 1&0\\
  0&1\endpmatrix\ {\rm mod}\ n\right\rbrace
$$
$$
  \Gamma_g(n,2n):=\left\lbrace M\in\Gamma_g(n)\, |\, {\rm
  diag}(A^tB)\equiv{\rm diag} (C^tD)\equiv0\ {\rm mod}\
  2n\right\rbrace.
$$
These are normal subgroups of $\Sp(2g,\Z)$ for $n>1$; however, $\Gamma_g(1,2)$ is not normal.
\end{df}
\begin{rem}
Theta constants with characteristics are not algebraically
independent, and satisfy a host of  algebraic identities. Their squares can be expressed algebraically in terms of a smaller set of modular forms, called theta constants of the second order, by using Riemann's bilinear addition theorem --- see \cite{I} for details. Theta constants of the second order are algebraically independent for $g\le 2$. The only identity among them for $g=3$ was known classically at least since the time of Schottky, and is discussed in \cite{vgvdg}, while the ideal of relations among them for $g>3$ is not known, and presumably very complicated.
\end{rem}

\section{The action of $\Sp(2g,\Z)$ on theta characteristics}
\begin{prop}[see \cite{I}]
Theta constants with characteristics are modular forms of weight one
half with respect to $\Gamma_g(4,8)$. Moreover, the full symplectic
group acts on theta constants with characteristics as follows:
$$
  \theta \bmatrix M\pmatrix \e\\ \de\endpmatrix\endbmatrix
  (M\cdot\tau)=\phi(\e,\,\de,\,M,\,
  \tau,\,z)\det(C\tau+D)^{\frac{1}{2}}\theta\bmatrix \e\cr \de
  \endbmatrix(\tau),
$$
where $\phi$ is some explicit eighth root of unity, and the action
on the characteristic is
\begin{equation}\label{actonchar}
  M\pmatrix \e\cr \de\endpmatrix :=\pmatrix D&-C\cr
  -B&A\endpmatrix\pmatrix \e\cr \de\endpmatrix+ \pmatrix {\rm diag}(C^tD)\\ {\rm diag}(A^tB)\endpmatrix
\end{equation}
where the addition in the right-hand-side is taken in $\F$.
\end{prop}
Notice that by this formula the action of the subgroup $\Gamma_g(2)$ on the set of characteristics is trivial, and thus the action of the entire group $\Sp(2g,\Z)$ on the set of characteristics facters through the action of $\Sp(2g,\Z)/\Gamma_g(2)=\Sp(2g,\F)$.

One can thus study the orbits of characteristics or sets of characteristics under the symplectic group action. This was done by Salvati Manni in \cite{SMlevel}, where all of the following results can be found. One first observes that the action of $\Gamma_g(4,8)\backslash\Gamma_g(2)$ on the set of theta constants differs from the modular one by extra signs, while $\Gamma_g(2)$ in addition permutes the characteristics. It is also clear that the action of $\Sp(2g,\Z)$ on the set of characteristics (which factors through the action of $\Sp(2g,\F)$) is transitive. To study the action on tuples of characteristics (i.e. the orbits of the $\Sp(2g,\F)$ acting on $\left(\F^{2g}\right)^n$ diagonally), we need more definitions.
\begin{df}
The Weil symplectic form on the space $\F^{2g}$ of characteristics is defined to be
$$
 \left\langle\tch\a\b,\tch\e\de\right\rangle:=\a\cdot\de+\b\cdot\e.
$$
Notice that this symplectic form is not preserved by the action of $\Sp(2g,\Z)$ on the set of characteristics: the pairing of the zero characteristic with any characteristic is zero, and it is the only such characteristic, while the action $\Sp(2g,\Z)$ given by (\ref{actonchar}) is affine and does not preserve zero.
\end{df}
\begin{df}
A triple of characteristics $\tch{\e_1}{\de_1},\tch{\e_2}{\de_2}, \tch{\e_3}{\de_3}$ is called syzygetic or azygetic depending on whether the sum
$$
  \e_1\cdot\de_1+\e_2\cdot\de_2+\e_3\cdot\de_3+(\e_1+\e_2+\e_3)\cdot
  (\de_1+\de_2+\de_3)
$$
$$
  = \left\langle\tch{\e_1}{\de_1},\tch{\e_2}{\de_2}\right\rangle+
  \left\langle\tch{\e_2}{\de_2},\tch{\e_3}{\de_3}\right\rangle +
  \left\langle\tch{\e_3}{\de_3},\tch{\e_1}{\de_1}\right\rangle
  \in \F
$$
is 0 or 1, respectively. Notice in particular that a triple of even
characteristics is syzygetic or azygetic if their sum is even or odd, respectively. This notion is in fact invariant under the symplectic group action.
\end{df}
The orbits of the action (\ref{actonchar}) of the symplectic group on sets of characteristics are completely described by the following
\begin{thm}[\cite{I} p.~212, \cite{SMlevel}]\label{action}
There exists an element of the symplectic group mapping a set of $n$
characteristics to another set of $n$ characteristics if and only if
there exists a way to number the characteristics in the first set
$a_1\ldots a_n$, and the characteristics in the second set $b_1\ldots b_n$ in such a way that
\begin{itemize}
\item for any $i$ the parity of $a_i$ and $b_i$ is the same
\item for any linear relation among $a_i$ with an even number of
    terms, i.e. if $a_{i_1}+\ldots+a_{i_{2k}}=0$, there is a
    corresponding linear relation $b_{i_1}+\ldots+b_{i_{2k}}=0$ and vice versa.
\item any triple $a_i, a_j, a_k$ is a/syzygetic if and only if
    the corresponding triple $b_i,b_j, b_k$ is a/syzygetic.
\end{itemize}
\end{thm}

\section{Results of Cacciatori, Dalla Piazza, van Geemen in terms of syzygy conditions}
The main new ingredient of the superstring measure in genus 3 proposed by Cacciatori, Dalla Piazza, and van Geemen in \cite{CDPvG} is the modular form $G$, of weight 8 with respect to the group $\Gamma(1,2)\subset \Sp(6,\Z)$ --- it is described there in terms of certain quadrics on $\F^6$. However, since $G$ is a polynomial in theta constants with characteristics, from theorem \ref{action} it follows that the monomials appearing in it should be characterized by the syzygy properties and linear dependencies of the characteristics involved. We now obtain such a description of the modular form $G$, by unraveling the definition of $G$ given in \cite{CDPvG} in terms of syzygies. We would like to thank Eric D'Hoker and Duong Phong for encouraging us to do this translation --- which then gave a formula amenable to generalizing to higher genus.

\smallskip
Given an even characteristic $\Delta=\tch{abc}{def}$ --- or $\tch\a\b$ in our notations --- one can define a corresponding quadratic form on the set of characteristics, i.e. (\cite{CDPvG}, p. 12) for $v\in\F^6$ one defines
$$
  q_\Delta(v):=v_1v_4+v_2v_5+v_3v_6+av_1+bv_2+cv_3+dv_4+ev_5+fv_6.
$$
If we write $v=\tch\e\de$, this is simply
$$
  q_\Delta(v)=\e\cdot\de+\a\cdot\e+\b\cdot\de,
$$
Since $\tch\a\b$ is an even characteristic, we have $\a\cdot\b=0$, and thus
$$
 q_\tch\a\b\left(\tch\e\de\right)=(\e+\b)\cdot(\de+\a).
$$
This looks strange as the $\a$ and $\b$ are strangely swapped, and I believe that there is a small typo in \cite{CDPvG} of interchanging the top and the bottom vector of the characteristic. For the characteristic $\tch{000}{000}$ that is used for explicit calculations in \cite{CDPvG} there is of course no difference, but otherwise modularity would not hold. The correct definition should thus be
\begin{equation}\label{defq}
 q_\tch\a\b\left(\tch\e\de\right)=(\e+\a)\cdot(\de+\b).
\end{equation}
In \cite{CDPvG} a quadric is now introduced
$$
  Q_\Delta:=\lbrace v\,|\, q_\Delta(v)=0\rbrace.
$$
From definition (\ref{defq}) it follows that this is the set of characteristics $\tch\e\de$ such that $\tch{\e+\a}{\de+\b}$ is even, i.e. this is just the set of even characteristics, to which $\tch\a\b$ is added. The symplectic form $\langle v,w\rangle$ on $\F^6$ is denoted $E(v,w)$ in \cite{CDPvG}. Notice that if both characteristics are even, this is the same as the quadratic form $q$. Considering Lagrangian (also called maximal isotropic classically, see \cite{I} and \cite{SM}) subspaces of $\F^6$ with respect to $\langle\cdot,\cdot\rangle$ means choosing three linearly independent characteristics $\tch{\e_i}{\de_i}_{i=1..3}$ such that the Weil pairing is zero on any pair, i.e. such that the sum of any pair of characteristics is again even. This is equivalent to saying that the triple of characteristics consisting of this pair and zero is syzygetic. The set of all even quadrics containing a Langrangian subspace is now considered in \cite{CDPvG}. This means considering the set of all characteristics $\Delta=\tch\a\b$ such that $q_\Delta|_L=0$. Thus
$$
 Q_\Delta\supset L\Longleftrightarrow \tch{\e+\a}{\de+\b}
 {\rm\ is\ even\ }\quad\forall\tch\e\de\in L.
$$
We now notice (following 8.4 in \cite{CDPvG}, essentially) that if $\Delta$ and $\Delta'$ are two characteristics such that $Q_\Delta\cap Q_{\Delta'}\supset L$, then we must have $\Delta-\Delta'\in L$. Thus the definition of $G$ at the top of page 13 in \cite{CDPvG} becomes
$$
  G[\Delta]=\sum\limits_{L\subset Q_\Delta}\prod\limits_{v\in L} \theta[v+\Delta]^2.
$$
The condition  $L\subset Q_\Delta$ means that the sum in the definition of $G$ is taken over linear spaces generated by triples of characteristics $\tch{\e_i}{\de_i}_{i=1..3}$ such that all $\tch{\a+\e_i}{\b+\de_i}$ are even and all $\left\langle \tch{\e_i}{\de_i}, \tch{\e_j}{\de_j}\right\rangle =\e_j\cdot\de_i+\e_i\cdot\de_j=0.$ Adding these conditions together shows that all characteristics $\tch{\a+\e_i+\e_j}{\b+\de_i+\de_j}=\tch\a\b+\tch{\a+\e_i}{\b+\de_i}
+\tch{\a+\e_j}{\b+\de_j}$ are even. Thus we get the following alternative formula
\begin{prop}
The modular form $G[\Delta]$ defined in \cite{CDPvG} (of weight 8 with respect to $\Gamma_3(1,2)$) is equal to the sum over all sets of 8 even characteristics $\lbrace u_i\rbrace_{i=1..8}$ such that any pair of characteristics together with $\Delta$ form a syzygetic triple, of the products $\prod\theta[u_i]^2$.
\end{prop}
Denote now $v_i:=u_i+\Delta$ and observe
$$
 \langle v_i,v_j\rangle=\langle u_i,u_j\rangle+\langle u_i,\Delta\rangle+\langle \Delta,u_j\rangle =0
$$
since the right-hand-side is exactly the condition that the triple $\Delta, u_i,u_j$ is syzygetic. We thus get yet another formula
\begin{cor}
The modular form $G$ can be written as
\begin{equation}\label{G3}
 G[\Delta]=\sum\limits_{V\subset\F^6\ \dim V=3}\ \prod\limits_{v\in V} \theta[v+\Delta]^2.
\end{equation}
\end{cor}
We remark that a given summand on the right-hand-side is not identically zero if and only if the set $V+\Delta$ contains only even characteristics (is a purely even coset in the language of \cite{SM}), in which case as described there it follows that $V$ is totally isotropic. This expression for $G$ yields itself to a straightforward generalization, and in this terms the restriction of $G$ to the locus of decomposable abelian varieties is easy to understand. We undertake this study in the next two sections.

\section{Modular forms corresponding to subspaces of $\F^{2g}$}
Motivated by the study of quartic relations among theta constants undertaken by Salvati Manni in \cite{SM} and by our reinterpretation of the form $G$ above, in this section we investigate the properties of products of theta constants with characteristics forming a translate of an isotropic subspace. We thank Riccardo Salvati Manni for telling us about \cite{SM} and encouraging us to explore the behavior of these modular forms.

Following \cite{SM}, we denote
$$
 P_M(\tau):=\prod\limits_{v\in M}\theta[v](\tau)\qquad
 {\rm for\ any}\quad M\subset\F^{2g}
$$
Notice that if $M$ contains any odd characteristics, then $P_M$ is identically equal to zero, as all odd theta constants vanish identically. Let $V\subset\F^{2g}$ be a vector subspace with basis $v_1\ldots v_n$. $V$ is called isotropic if the symplectic form restricts to zero on it, i.e. if $\langle v,w\rangle=0$ for any $v,w\in V$. Since for even characteristics $v$ and $w$ the value of the symplectic form $\langle v,w\rangle$ is equal to the parity of $v+w$, the space with basis $\lbrace v_i\rbrace$ is isotropic if and only if it only contains even characteristics, or equivalently, if $P_V(\tau)$ is not identically zero.
\begin{df}
We define a function $P_{i,s}^{(g)}$ on $\H_g$ as the sum
$$
  P_{i,s}^{(g)}(\tau):=\sum\limits_{V\subset\F^{2g};\, \dim V=i} P_V(\tau)^s.
$$
Here the sum is taken over all $i$-dimensional (and thus of cardinality $2^i$) vector subspaces, but note that if $V$ is not isotropic, it contains odd characteristics, and the corresponding summand is zero. {\it We will only consider these functions subject to the condition $2^is=2^k$ for some integer $k\ge 4$. For the case of the superstring measure $k=4$.}
\end{df}
\begin{prop}\label{Pismodular}
The function $P_{i,s}^{(g)}$ is a modular form (assuming $2^is=2^k$ for $k\ge 4$) of weight $2^{i-1}s$ for the subgroup $\Gamma_g(1,2)$.
\end{prop}
\begin{proof}
This can be seen from the discussion in \cite{SM}, \cite{I}, \cite{K}. To see this, one notes that the action of $\Sp(2g,\Z)$ on the set of characteristics is affine; however, for an element $\gamma=\pmatrix A&B\\ C&D\endpmatrix\in \Gamma_g(1,2)$ by definition ${\rm diag}(C^tD)= {\rm diag}(A^tB)=0\in\F^g$, and thus the action of $\Gamma_g(1,2)$ on the set of characteristics fixes zero and is linear. Thus the summands in the definition of $P_{i,s}$ get permuted --- vector subspaces are mapped to vector subspaces by a linear action. The multiplicative factors are $\det(C\tau+D)^{1/2}$ for each theta constant, giving the overall factor of $\det(C\tau+D)^{2^{i-1}}$ for each $P_V$. The other factor in the transformation formula (\ref{actonchar}) is the 8th root of unity $\phi$. Since $2^is=2^k$ is divisible by 16, it can be shown that the product of the 8th roots will turn out to be equal to 1. We refer to \cite{SMnew} for a complete discussion and rigorous proof.
\end{proof}
The action of the full group $\Sp(2g,\Z)$ on the set of characteristics is affine --- it shifts the zero to some characteristic $\Delta$. Acting by $\Sp(2g,\Z)$ on the modular form $P_{i,s}^{(g)}$ we then get
\begin{cor}
Assuming $2^is=2^k$ for $k\ge 4$, for any even characteristic $\Delta$ the function
$$
  P_{i,s}^{(g)}[\Delta](\tau):=\sum\limits_{V\subset\F^{2g};\,\dim V=i} P_{V+\Delta}(\tau)^s
$$
is a modular form of weight $2^{i-1}s$ with respect to the subgroup $\Gamma[\Delta]\subset\Sp(2g,\Z)$ that stabilizes $\Delta$ under the action (\ref{actonchar}). This subgroup is conjugate to $\Gamma_g(1,2)$ (which, recall, is not a normal subgroup of $\Sp(2g,\Z)$). The conjugation is provided by any element of $\Sp(2g,\Z)$ that maps the zero characteristic to $\Delta$ under the action (\ref{actonchar}). Note that if $\Delta$ is odd, the corresponding expression would be zero, as all summands would contain $\theta[\Delta]$.
\end{cor}
\begin{proof}
To prove modularity, one can observe that $P_{i,s}^{(g)}=P_{i,s}^{(g)}[0]$, which we know to be a modular form, is conjugated to $P_{i,s}^{(g)}[\Delta]$ by the $\Sp(2g,\Z)$ action. For a direct proof, note that $v+\Delta$ is an even characteristic for all $v\in V$ if and only if for any triple $v_1,v_2,\Delta$ for $v_1,v_2\in V$ is syzygetic --- this is the argument used to obtain the expression (\ref{G3}) in genus 3 at the end of the previous section --- and the syzygy is preserved by the $\Sp(2g,\Z)$ action. See \cite{SMnew} for more discussion, especially on the possible 8th roots of unity.
\end{proof}

\section{Restrictions of modular forms corresponding }
In this section we determine the restrictions of modular forms $P_{i,s}^{(g)}$ to the loci of decomposable abelian varieties (products of lower-dimensional ones). Note that it is enough to determine the restriction of the modular form $P_{i,s}^{(g)}$ with zero characteristic to $\H_k\times \H_{g-k}$ --- its restrictions to the conjugates of this locus under $\Gamma_g(1,2)$ can be obtained by acting by $\Gamma_g(1,2)$, which preserves the form. Furthermore, the restrictions of $P_{i,s}^{(g)}[\Delta]$ with non-zero characteristic can then be obtained by  conjugating by an element of $\Sp(2g,\Z)$ that maps $P_{i,s}^{(g)}$ to $P_{i,s}^{(g)}[\Delta]$.
\begin{thm}\label{restrict}
The modular form $P_{i,s}^{(g)}$ restricts to the locus of decomposable abelian varieties (reducible period matrices in \cite{CDPvG}) as follows:
\begin{equation}
 P_{i,s}^{(g)}|_{\H_k\times\H_{g-k}}= \sum\limits_{0\le n,m\le i\le n+m} N_{n,m;\, i\,}P_{n,2^{i-n}s}^{(k)}\cdot P_{m,2^{i-m}s}^{(g-k)},
\end{equation}
where
\begin{equation}\label{Ncoeff}
 N_{n,m;\, i\,} =\prod\limits_{j=0}^{n+m-i-1}\frac{(2^n-2^j)(2^m-2^j)}{
 (2^{n+m-i}-2^j)}.
\end{equation}
\end{thm}
\begin{rem}
Notice that many of the summands can be zero, as $P_{a,b}^{(c)}\equiv 0$ for $a>c$. In particular for $P_{g,s}^{(g)}$ (the case of maximal isotropic subspaces; the form $G$ constructed in \cite{CDPvG} is $P_{3,2}^{(3)}$ in our notations) the only non-zero term on the right is $P_{k,2^{g-k}s}^{(k)} P_{g-k,2^ks}^{(g-k)}$.
\end{rem}
\begin{proof}
Indeed, let $V\cong \F^i$ be a vector subspace of $\F^{2g}$ (notice that we never need to worry about parity or isotropy --- the summands for non-isotropic subspaces will vanish automatically). If a period matrix is a product of two lower-dimensional ones, $\tau_g=\tau_k\times \tau_{g-k}$, the group of points of order two on $A_\tau$ is the direct sum of the groups of points of order two on the factors. Let $\F^{2g}=\F^{2k}\oplus\F^{2(g-k)}$ be this decomposition, and let $\pi_1$ and $\pi_2$ denote the projections onto the two summands. Since $V\subseteq \pi_1(V)\oplus\pi_2(V)$, we must have $\#V\le \#\pi_1(V)\cdot\#\pi_2(V)$. Since these are all vector spaces over $\F$, denoting by $n,m$ the dimensions of $\pi_1(V)$ and $\pi_2(V)$ respectively, this implies the inequality $2^i\le 2^n\cdot 2^m$ or, equivalently, $i\le n+m$. The projections maps $\pi_1:V\to\pi_1(V)$ and $\pi_2:V\to\pi_2(V)$, being group homomorphisms, are then $2^{i-n}$-to-1 and $2^{i-m}$-to-1, respectively. Since any theta constant with characteristic restricts to the product
$$
 \theta[v](\tau_g)=\theta[\pi_1(v)](\tau_k)\cdot \theta[\pi_2(v)](\tau_{g-k}),
$$
it follows that for $n$ and $m$ fixed we have
$$
 P_V(\tau_g)^s=\prod\limits_{v\in V}\theta[v](\tau_g)^s
 =\prod\limits_{v\in V}\theta[\pi_1(v)](\tau_k)^s\cdot \theta[\pi_2(v)](\tau_{g-k})^s
$$
$$
 =
 \prod\limits_{v_1\in\pi_1(V)}\theta[v_1](\tau_k)^{2^{i-n}s}
 \prod\limits_{v_2\in\pi_2(V)}\theta[v_2](\tau_{g-k})^{2^{i-m}s}
 =P_{\pi_1(V)}^{2^{i-n}s}\cdot P_{\pi_2(V)}^{2^{i-m}s}
$$
To compute $P_{i,s}^{(g)}$, we need to sum over all $V$. Let us first sum over all the spaces $V$ for which the spaces $\pi_1(V)$ and $\pi_2(V)$ are fixed. Notice that the product on the right is the same for all $V$ with fixed projections. The number of $V$ with fixed projections only depends on the dimensions, and we compute it in the following combinatorial lemma.
\begin{lm}\label{comblm}
The number of $i$-dimensional vector subspaces $V$ of $\F^n\oplus\F^m$ surjecting onto both summands
$$
 \prod\limits_{j=0}^{n+m-i-1}\frac{(2^n-2^j)(2^m-2^j)}{
 (2^{n+m-i}-2^j)},
$$
(where this number is understood to be zero if $n+m<i$, and to be one if $n+m=i$. Note also that the product has a zero factor if $n>i$ or $m>i$).
\end{lm}
\begin{proof}
Fix a scalar product on $\F^{n+m}$ such that the chosen decomposition $\F^{n+m}=\F^n\oplus \F^m$ is orthogonal. The projections from $V$ to $\F^n$ and $\F^m$ are then the orthogonal projections; the image of such a projection misses a vector $v$ if and only if it is orthogonal to $V$. Thus what we need to count is (for a fixed scalar product) the number of $V\cong\F^i$ such that $V^\perp\cong\F^{n+m-i}$ does not intersect the coordinate subspaces $\F^n$ and $\F^m$ away from zero. Let us construct such a $V^\perp$ by choosing a basis $v_1,\ldots, v_{n+m-i}$ for it. To choose such a basis, we will choose independently the projections $\pi_1(v_1),\ldots,\pi_1(v_{n+m-i})$ --- note that in order to have $V^\perp\cap\F^n=\lbrace 0\rbrace$, these vectors must be linearly independent --- and similarly choosing linearly independent $\pi_2(v_1),\ldots,\pi_2(v_{n+m-i})\in\F^m$.
Thus $\pi_1(v_1)$ can be chosen in $2^n-1$ ways, after which $\pi_1(v_2)$ can be chosen in $2^n-2$ ways, and so on, and similarly we choose $\pi_2(v_1)$ in $2^m-1$ ways and so on. Thus the total number of spaces $V^\perp\subset\F^{n+m}$ not intersecting $\F^n$ and $\F^m$, together with a choice of an ordered basis of it, is equal to
$$
 \prod\limits_{j=0}^{n+m-i-1}(2^n-2^j)(2^m-2^j)
$$
while the number of ordered basis in a fixed space $V^\perp\cong\F^{n+m-i}$ is
$$
  \prod\limits_{j=0}^{n+m-i-1}(2^{n+m-i}-2^j),
$$
and dividing one by the other gives the lemma.
\end{proof}
Now observe that if we take the sum over all $V$ for fixed dimensions $n$ and $m$, the projections $\pi_1(V)$ and $\pi_2(V)$ range over all $n$-dimensional subspaces of $\F^{2k}$, and $m$-dimensional subspaces of $\F^{2(g-k)}$, respectively. We thus get
$$
 \sum\limits_{V\subset\F^{2g};\ \dim V=i;\ \dim \pi_1(V)=n;\ \dim \pi_2(V)=m}P_V(\tau_g)^s
$$
$$
 =N_{n,m;\,i\,}\left(\sum\limits_{V_1\subset\F^{2k}\dim V_1=n}P_{V_1}^{2^{i-n}s}\right)\cdot
\left(\sum\limits_{V_2\subset\F^{2(g-k)}\dim V_2=m}P_{V_2}^{2^{i-m}s}\right)
$$
$$
 =N_{n,m;\,i\,}P_{n,2^{i-n}s}^{(k)}\cdot P_{m,2^{i-m}s}^{(g-k)}
$$
This is of course zero if $n>k$ or $m>g-k$. Taking the sum over all possible $n$ and $m$ (recall that we have $n+m\ge i$ and $n,m\le i$) gives the theorem.
\end{proof}

\section{Ans\"atze for genera $\le 4$ in terms of vector subspaces}
We now rewrite the low genus superstring measure proposed by D'Hoker and Phong \cite{DHP1} for genus 2 and by Cacciatori, Dalla Piazza, and van Geemen \cite{CDPvG} for genus 3 in terms of the modular forms $P_{i,s}^{(g)}$ constructed and studied above.
Following the earlier works, we try to find the superstring measure as the product of the bosonic measure (some formulas for which are known, but explicit point-independent expressions for which are apparently not known for high genus --- see the discussion in \cite{DHP1}) and a function $\Xi^{(g)}[\Delta]$ depending on the characteristic $\Delta$. As argued in \cite{CDPvG}, p.5 the factorization property can then be rewritten as a condition on $\Xi^{(g)}[\Delta]$. The arguments in \cite{CDPvG}, 2.7 show (essentially arguing that if $\Xi^{(g)}[\Delta]$ is equal to $\gamma\Xi^{(g)}$ for some $\gamma\in\Sp(2g,\Z)$, then its degenerations can be computed by acting by $\gamma$ on the degenerations of $\Xi^{(g)}$) that for $g\le 3$ to satisfy these constraints it is enough to construct a holomorphic modular form $\Xi^{(g)}$ of weight 8 with respect to $\Gamma_g(1,2)$ satisfying the factorization constraint
$$
 \Xi^{(g)}|_{\H_k\times\H_{g-k}}=\Xi^{(k)}\cdot \Xi^{(g-k)}
$$
for any $k<g$. The reason this statement is only proven for $g\le 3$ is that the superstring measure a priori may only be defined on the moduli space of curves $\M_g$ and not on the entire space $\A_g$. For genus high enough the locus of decomposable abelian varieties $\A_k\times \A_{g-k}$ does not lie in (the closure of) the locus of Jacobians, and the superstring measure on $\M_g$ may not give rise to a modular form on $\A_g$. However, for genus 4 any decomposable abelian variety is still a product of Jacobians (possibly of nodal curves), and the statement still holds.

\smallskip
We will now rewrite the known low genus superstring measures in terms of our forms $P_{i,s}^{(g)}$. Notice that we are looking for a form $\Xi^{(g)}$ of weight 8, and thus we will look for it as a linear combination of
\begin{equation}\label{defF}
  \P{i}{g}:=P_{i,2^{4-i}}^{(g)},
\end{equation}
which are the only $G's$ of appropriate weight. Since the only $0$-dimensional vector space is zero, and all the $1$-dimensional spaces over $\F$ consist of zero and another vector, we have
$$
 \P0g=\theta[0]^{16};\qquad \P1g=\theta[0]^8\sum\limits_{v\in\F^{2g}\setminus\lbrace 0\rbrace}\theta[v]^8.
$$
In genus 1 the superstring measure is known to be given by
$$
 \Xi^{(1)}=\tt00^4\eta^{12}=\tt00^8\tt01^4\tt10^4.
$$
The modular forms we have are
\begin{equation}\label{F01}
  \P01=\tt00^{16}\qquad{\rm and}\qquad \P11=\tt00^8\left(\tt01^8+\tt10^8\right)
\end{equation}
Using the Jacobi relation --- the only algebraic identity among the three even theta constants with characteristics in genus 1 --- we can express $\Xi^{(1)}$ in terms of these, obtaining
$$
  \Xi^{(1)}=\frac12\left(\P01-\P11\right).
$$
In fact looking in \cite{CDPvG}, p.9 we see that
$$
 \P01=\tt00^4\left(\frac13 f_{21}+\eta^{12}\right)\qquad{\rm and} \qquad \P11=\tt00^4\left(\frac13 f_{21}+\eta^{12}\right).
$$

\smallskip
The genus 2 superstring measure was computed in \cite{DHP1}, but us let us try to look for it in the form
$$
 \Xi^{(2)}=a_0\P02+a_1\P12+a_2\P22.
$$
The decomposable locus in this case is $\H_1\times\H_1$ (together with its $\Sp(4,\F)$ conjugates), the restriction of $\Xi^{(2)}$ by theorem \ref{restrict} is
$$
 a_0\P01\cdot\P01+a_1\left(\P01\cdot\P11+\P11\cdot\P01+\P11\cdot\P11\right)
 +a_2\P11\cdot\P11.
$$
Notice that here all the combinatorial coefficients $N_{n,m;\,i\,}$ from theorem \ref{restrict} are equal to one, which is easy to see geometrically by remembering that this is the count of the number of $i$-dimensional subspaces of $\F^n\oplus\F^m$ projecting onto both factors. Some of the coefficients for the restrictions we compute for higher genus are not as obvious, and we make substantial use of the theorem. For this restriction to be equal to
$$
 \Xi^{(1)}\cdot\Xi^{(1)}
 =\frac14\P01\cdot\P01-\frac14\P01\P11-\frac14\P11\cdot\P01 +\frac14 \P11\cdot\P11
$$
we must choose $a_0=\frac14, a_1=-\frac14, a_2=\frac12$, thus verifying
\begin{prop}
The following ansatz:
$$
  \Xi^{(2)}=\frac14\left(\P02-\P12+2\P22\right),
$$
while being a holomorphic modular form of weight 8 with respect to $\Gamma_2(1,2)$, satisfies the factorization constraint.
\end{prop}
\begin{rem}
Note that unlike the formula for the genus 2 amplitude obtained in \cite{DHP1}, and the various expressions for it studied in \cite{DHP2}, our formula for $\Xi^{(2)}$ involves syzygetic rather than azygetic sets --- and they all include the zero characteristic, so that the $\theta[0]^4$ factors appears naturally. To show that our ansatz is equal to the formulas given in \cite{DHP1} and \cite{CDPvG} one expresses all theta functions with characteristics in terms of theta functions of the second order using the bilinear addition theorem and verifies that an identity is obtained (using Maple, not by hand).
\end{rem}
We now search for a genus 3 ansatz in the form
$$
 \Xi^{(3)}=a_0\P03+a_1\P13+a_2\P23+a_3\P33.
$$
Notice that this form is equivalent to the one used in \cite{CDPvG}: our $\P33$ is their $G$, and their $F_i$'s can be expressed as linear combinations of our $\P03,\P13,\P23$, as can be verified by implementing the bilinear addition theorem in Maple. In our form we can use theorem \ref{restrict} to easily compute the restriction of $\Xi^{(3)}$ to the locus of decomposable abelian varieties, which again has only one component, $\H_1\times\H_2$, to be
$$
 \Xi^{(3)}|_{\H_1\times\H_2}=a_0\P01\cdot\P02
 +a_1\left(\P01\cdot\P12+\P11\cdot\P02+\P11\cdot\P12\right)
$$
$$ +a_2\left(\P01\cdot\P22+\P11\cdot\P12+3\P11\cdot\P22\right)
 +a_3\P11\cdot\P22.
$$
Requiring this to be equal to
$$
 \Xi^{(1)}\cdot\Xi^{(2)}=\frac18\left(\P01-\P11\right)
 \left(\P02-\P12+2\P22\right)
$$
$$
=\frac18\left(\P01\cdot\P02-\P01\cdot\P12-\P11\cdot\P02\right.
$$
$$
\left.+2\P01\cdot\P22+\P11\cdot\P12-2\P11\cdot\P22\right)
$$
(we arranged the terms to be in the same order as in the formula for the restriction, where we note that $\P11\cdot\P12$ and $\P11\cdot\P22$ appear twice) allows us to compute the coefficients $a_i$ starting from $i=0$ uniquely to get
\begin{prop}
The following ansatz:
$$
  \Xi^{(3)}=\frac18\left(\P03-\P13+2\P23-8\P33\right),
$$
while being a holomorphic modular form of weight 8 with respect to $\Gamma_3(1,2)$, satisfies the factorization constraints.
\end{prop}

\bigskip
We now look for a genus 4 ansatz in the form
$$
 \Xi^{(4)}=a_0\P04+a_1\P14+a_2\P24+a_3\P34+a_4\P44.
$$
This is the first case when we have two different reducible loci: $\H_1\times\H_3$ and $\H_2\times\H_2$. We again use theorem \ref{restrict} to compute the restrictions. To unclutter the formulas we drop the upper indices on the $P$'s and compute
$$
 \Xi^{(4)}|_{\H_1\times\H_3}=a_0 G_0\cdot G_0
 +a_1(G_0\cdot G_1+G_1\cdot G_0+G_1\cdot G_1)
$$
$$
 +a_2(G_0\cdot G_2+G_1\cdot G_1+3G_1\cdot G_2)
 +a_3(G_0\cdot G_3+G_1\cdot G_2+7G_1\cdot G_3)
 +a_4 G_1\cdot G_3
$$
Requiring this to be equal
$$
 \Xi^{(1)}\cdot\Xi^{(3)}=\frac{1}{16}(G_0-G_1)\cdot(G_0-G_1+2G_2-8G_3)
$$
we can solve for $a_i$ term by term (and the solution is unique!). On the other hand, we must also have a correct factorization on $\H_2\times\H_2$: we must have
$$
 \Xi^{(4)}|_{\H_2\times\H_2}=a_0G_0\cdot G_0
 +a_1(G_0\cdot G_1+G_1\cdot G_0+G_1\cdot G_1)
$$
$$
 +a_2(G_0\cdot G_2+G_1\cdot G_1+3G_1\cdot G_2+G_2\cdot G_0+3G_2\cdot G_1+ 6G_2\cdot G_2)
$$
$$
 +a_3(G_1\cdot G_2+G_2\cdot G_1+ 9G_2\cdot G_2)
 +a_4 G_2\cdot G_2
$$
equal to
$$
  \Xi^{(2)}\cdot\Xi^{(2)}=\frac{1}{16}(G_0-G_1+2G_2)\cdot(G_0-G_1+2G_2).
$$
This can again be solved term by term to give a unique solution for all $a_i$. Miraculously these solutions are the same (this is best checked by Maple, or, very tediously, by hand)) and we thus get
\begin{thm}
The expression
$$
  \Xi^{(4)}:=\frac{1}{16}\left(\P04-\P14+2\P24-8\P34+64\P44\right)
$$
is a modular form of weight 8 with respect to $\Gamma_4(1,2)$, and satisfies the factorization constraints. This is the unique such linear combination of $\P{i}{4}$, and is thus a natural candidate for the genus 4 superstring measure.
\end{thm}

\section{Further directions}
There seem to be three natural further questions to ask, which we now discuss one by one.

\subsubsection*{Question 1: Propose an ansatz for the superstring measure in any genus}

The computations above seem to work miraculously, but this is of course not a coincidence. By working carefully with the combinatorics of the coefficients in the restriction formula in theorem \ref{restrict} one can always get a unique linear combination of $\P{i}{g}$ that restricts correctly.
\begin{thm}\label{Thmg}
For any genus $g$ the (possibly multivalued) function
$$
 \Xi^{(g)}:=\frac{1}{2^g}\sum\limits_{i=0}^g (-1)^i2^{\frac{i(i-1)}{2}}\P{i}{g}
$$
is a modular form of weight 8 (up to a possible inconsistency in the choice of roots of unity for the different summands) with respect to $\Gamma_g(1,2)$, such that its restriction to $\H_k\times\H_{g-k}$ is equal to $\Xi^{(k)}\cdot\Xi^{(g-k)}$.

Moreover, $\Xi^{(g)}$ is the unique linear combination of $\P{i}{g}$ that restricts to the decomposable locus in this way.
\end{thm}
\begin{proof}
Uniqueness is easy to see: indeed, the coefficients of $\P{i}{g}$ can be computed inductively; notice that the coefficient of $G_i$ has to be the same for all genera, as there is no dependence on $g$ in the formula for restrictions --- see (\ref{induct}) below. Thus in genus $g$ there is in fact only one new coefficient to compute, that in front of $\P{g}{g}$, and this can be computed from the coefficient of $\P{1}{1}\cdot\P{g-1}{g-1}$ when restricting to $\H_1\times\H_{g-1}$.

The hard part is verifying that this ansatz works --- a priori it could happen that imposing the restriction constraint for some $\H_k\times\H_{g-k}$ would be incompatible with the constraint for some other $k'$. Thus we need to verify that for the ansatz above we have for all $k$
$$
 \Xi^{(g)}|_{\H_k\times\H_{g-k}}=\Xi^{(k)}\cdot\Xi^{(g-k)}.
$$
Notice that the product on the right-hand-side is a sum of products of the type $\P{n}{k}\cdot\P{m}{g-k}$ with coefficients given by the lower-genus ans\"atze. Theorem \ref{restrict} shows that the left-hand-side is also a sum of terms of the same kind, and thus what we need to prove is that the coefficients of $\P{n}{k}\cdot\P{m}{g-k}$ on both sides agree, i.e. that (notice that the powers of $\frac12$ all cancel)
\begin{equation}\label{induct}
 \sum\limits_{i=0}^{n+m}(-1)^i2^{\frac{i(i-1)}{2}}N_{n,m;\, i\,} =(-1)^n2^{\frac{n(n-1)}{2}}\cdot(-1)^m2^{\frac{m(m-1)}{2}},
\end{equation}
where $N_{n,m;\,i\,}$ is the number of $i$-dimensional subspaces of $\F^n\oplus\F^m$ surjecting onto both summands, given explicitly by (\ref{Ncoeff}). Note that the summands for $i<\max(n,m)$ are automatically zero, but it is convenient to include them formally. Notice that this identity does not depend on $g$ and $k$, which is why the coefficient of $\P{i}{g}$ in $\Xi^{(g)}$ does not depend on $i$.

While the quantity $N$ has a geometric interpretation and thus (\ref{induct}) seems amenable to a geometric inclusion-exclusion proof, we give an easy proof by induction, still using some geometry for the inductive step. Note that $n$ and $m$ enter the formula symmetrically, so we can induct in either, and note that identity (\ref{induct}) is obviously true for $n=0$ or $m=0$, when there is only one summand on the left, and there is no product to take. To perform induction, we use the following combinatorial
\begin{lm}
The counting functions $N_{n,m:\,i\,}$ given explicitly by (\ref{Ncoeff}) satisfy the following recursion:
\begin{equation}\label{recur}
  N_{n,m+1;\,i+1\,}=N_{n,m;\, i\,}+(2^{i+1}-2^m)N_{n,m;\, i+1\,}.
\end{equation}
\end{lm}
\begin{proof}
Recall that $N_{n,m+1;\,i+1\,}$ is the number of $(i+1)$-dimensional subspaces $V$ of $\F^n\oplus\F^{m+1}$ surjecting onto both summands under the projection maps $\pi_1$ and $\pi_2$. Let $p:\F^n\oplus\F^{m+1}\to\F^n\oplus\F^m$ be the projection forgetting the last basis vector (denote this vector by $e_{m+1}$). Since $p$  is a homomorphism, the map $p:V\to p(V)$ is either 2-to-1, in which case $\dim p(V)=i$ and $V=p^{-1}(V)$, or it is 1-to-1 and $\dim p(V)=i+1$. We now count how many different $V$ can give rise to a given $p(V)\subset\F^n\oplus\F^m$ (which still surjects onto both summands). If $\dim p(V)=i$, then $V=p^{-1}(V)$ is unique --- thus we get the first summand in the lemma, with no coefficient.

For the second case, choose $(p_1,\ldots,p_m)\in p(V)$ such that $p\circ\pi_2(p_k)=e_k$ is the $k$'th basis vector of $\F^m$ (this is possible since $p\circ\pi_2:V\twoheadrightarrow\F^m$). The vectors $\lbrace p_1,\ldots, p_m\rbrace\in p(V)$ are linearly independent since their projections are. We can complete them to a basis $\lbrace p_1, \ldots, p_{i+1}\rbrace$ of $p(V)$ such that $p\circ\pi_2(p_k)=0$ for $m<k\le i+1$: to accomplish this, take any basis of $p(V)$ and subtract the appropriate sums of $p_1\ldots p_m$ from the rest to make $p\circ\pi_2$ zero.

To determine $V$ given this $p(V)$ it suffices to choose $v_1,\ldots, v_{i+1}$ such that $p(v_k)=p_k$, which amounts to choosing the $e_{m+1}$-coordinate of each $v_k$ --- thus there are $2^{i+1}$ choices. For any such choice $V$ will surject onto $\F^n$ and onto $\F^m$, but for $V$ to surject onto $\F^{m+1}$ it is necessary and sufficient for there to exist $m+1$ vectors in $V$ with linearly independent $\pi_2$ projections. We can choose $v_1,\ldots, v_m$ as $m$ of this vectors, and thus the condition for $\pi_2:V\to\F^{m+1}$ to be surjective is for there to exist some vector in the span of $v_{m+1},\ldots,v_{i+1}$ with non-zero $e_{m+1}$-coordinate. Thus unless the $e_{m+1}$ coordinate of all $v_k$ is zero, i.e. unless $v_k=p_k$ for all $k=m+1,\ldots,i$ (and then we have two choices for each of $v_1,\ldots,v_m$ --- so there are $2^m$ such cases), $V$ surjects onto both summands of $\F^n\oplus\F^{m+1}$ and is counted in $N_{n,m+1;\,i+1\,}$. Thus for each $p(V)$ of dimension $i+1$ there are exactly $2^{i+1}-2^m$ different subspaces $V$ projecting to it that are counted in $N_{n,m+1;\,i+1\;}$.
\end{proof}
\begin{rem}
Note that this proof works also in the case when some of the $N$'s appearing in the formula are zero. Another proof of the lemma can be obtained (but not so easily guessed!) by writing out the formulas (\ref{Ncoeff}) for $N$'s in terms of products and manipulating them using $2^{k+1}-2^{j+1}=2(2^k-2^j)$, etc. Then one also has to check the cases when some $N$ is zero separately, while the geometric argument works in all cases.
\end{rem}
We now complete the proof of the theorem by inducting from $m$ to $m+1$. We substitute the recursive expression (\ref{recur}) from the lemma into the left-hand-side of (\ref{induct}) to get (we use $I=i+1$ for the index of summation)
$$
  \sum\limits_{I=0}^{n+m}(-1)^I2^{\frac{I(I-1)}{2}}N_{n,m+1;\, I\,}
$$
$$ =  \sum\limits_{I=0}^{n+m+1}(-1)^I2^{\frac{I(I-1)}{2}}\left(
  N_{n,m;\, I-1\,}+(2^I-2^m)N_{n,m;\, I\,}\right)
$$
$$
  =-2^m\sum\limits_{I=0}^{n+m}(-1)^I2^{\frac{I(I-1)}{2}}N_{n,m;\,I\,}
$$
$$
  +\sum\limits_{i=0}^{n+m}(-1)^{i+1}2^{\frac{(i+1)i}{2}}N_{n,m;\,i\,}
  +\sum\limits_{I=0}^{n+m}(-1)^I2^{\frac{I(I-1)}{2}}2^IN_{n,m;\,I\,}
$$
where we used the fact that the $i=-1$ and $I=n+m+1$ summands are zero in the last two sums, and the fact that formula (\ref{recur}) works for $N$'s some of which are zero as well. Now we note that the two expressions in the last line are the same up to sign and renaming the variable from $i$ to $I$, and thus they cancel, so that we finally use the inductive assumption that (\ref{induct}) holds for $m$ to obtain
$$
 -2^m\sum\limits_{I=0}^{n+m}(-1)^I2^{\frac{I(I-1)}{2}}N_{n,m;\,I\,}
 =-2^m(-1)^{n+m}2^{\frac{n(n-1)}{2}+\frac{m(m-1)}{2}}
$$
which is equal to the expression
$$
 (-1)^{n+m+1}2^{\frac{n(n-1)}{2}+\frac{m(m+1)}{2}},
$$
the right-hand-side of (\ref{induct}) for $n$ and $m+1$. The step of the induction is thus proven.
\end{proof}
\begin{rem}
This ansatz is a direct generalization of the formulas we obtained above for $g\le 4$ (and of course agrees with those).
The potential problem with the multivaluedness here stems from the fact that for example $G_5=P_{5,\frac12}$ is the sum of square roots of products of thetas. It could well happen, and seems perhaps not quite unlikely in view of Riemann's quartic relations and Schottky-Jung identities (for an example of the Riemann quartic relation and the identities for theta constants on the Schottky locus, see the discussion of the genus 3 situation in \cite{vgvdg}), that the product of 32 theta constants with characteristics in a vector subspace may indeed admit a holomorphic root over $\M_g$. In this case the expression above would be a natural candidate for the superstring measure. Note that the product of all theta constants has a holomorphic square root in genus 3 by results of Igusa; this kind of condition for the square root to be holomorphic was also encountered in the first attempts to compute the genus 3 superstring measure in \cite{DHP2,DHP3}.
\end{rem}

\subsubsection*{Question 2: Verify that the proposed ansatz satisfies further physical constraints, for example that it yields a vanishing cosmological constant and vanishing 2- and 3-point functions}

Showing that the cosmological constant vanishes is equivalent to showing that the sum $\sum_\Delta\Xi^{(g)}[\Delta]$ is identically zero. This has been verified for genus 2 in \cite{DHP1} and  for genus 3 in \cite{CDPvG}. In general notice that this sum, if non-zero, is a modular form with respect to the entire group $\Sp(2g,\Z)$ of weight 8. From the factorization constraint being satisfied we know that it restricts to the locus of decomposable abelian varieties as the product of the corresponding lower-dimensional sums, which we can inductively assume to vanish. In particular in genus 4 this sum vanishes on the locus $\A_3\times\A_1$ and thus on the boundary of $\A_g$, which implies that this sum, if non-zero, is a modular form of slope at most 8.

However, it is known that the slope of the effective cone of $\M_4$ is equal to $6+\frac{12}{4+1}>8$, and it is in fact known that the Schottky locus $\M_4\subset\A_4$ is the zero locus of the unique modular form of slope 8 on $\A_4$, the Schottky equation. Thus the form $\sum_\Delta\Xi^{(4)}[\Delta]$ must be a (possibly zero) multiple of this Schottky equation, and thus vanishes identically on $\M_4$, so our ansatz does produce a vanishing cosmological constant in genus 4.

It seems very hard to extend a similar kind of argument to higher genus, where the slopes of effective divisors on $\M_g$ and $\A_g$ are not known.

Another constraint on the measure is to verify that all the 2- and 3-point functions vanish. This was verified for the genus 2 measure in \cite{DHPd,DHPe}, and checking this for the proposed ans\"atze in genera 3 and 4 would be a good indication of their potential validity.

\subsubsection*{Question 3: Investigate whether the above restrictions are sufficient to guarantee the uniqueness of the solution for the superstring measure}

If we restrict ourselves to looking for the superstring measure as a product of the bosonic measure and a modular form of weight 8, suppose $g$ is the lowest genus for which the ansatz is not unique, i.e. when there exist two distinct modular forms of weight 8 for $\Gamma_g(1,2)$ with the identical restriction to the decomposable locus. Then their difference $F$ would be a modular form $F$ of weight 8 with respect to $\Gamma_g(1,2)$ vanishing on all the components $\A_k\times\A_{g-k}$ of the locus of decomposable abelian varieties in $\A_g$. If it could be shown from the theory of modular forms that such an $F$ is then identically zero, then uniqueness of the ansatz in genus $g$ would follow. This seems to be a really hard question, as the ring of modular forms for $\Gamma_g(1,2)$ for genus $g\ge 4$ is not generated by theta constants --- see the recent results in \cite{OSM}, \cite{OPY}. In general describing the ring of modular forms for $\Gamma_g(1,2)$ and obtaining conditions guaranteeing the vanishing of such a form seems very hard --- see \cite{PY}. The authors of \cite{CDPvG} indicate that they will give a proof of uniqueness in genus 3 in a forthcoming paper.

\smallskip
Also note that while in \cite{DHP1,DHPa,DHPb,DHPc} the formula for the genus 2 superstring measure was derived from the first principles and as such has to be unique, the derivations in higher genus are based on the assumption of the superstring measure being a product of a modular form and the bosonic measure, which then needs to be justified in some physical way.

\section*{Acknowledgements}
I learned about the problem from Duong Phong, to whom I am very grateful for his constant encouragement, for explaining the basic questions and computations for the string measure, and for detailed comments on the draft of this text. I am very thankful to Eric D'Hoker and Duong Phong for many conversations on the subject, for sharing their Maple code, for ideas on how the higher genus superstring measure could be constructed, and especially for suggesting that I express the genus 3 ansatz of Cacciatori, Dalla Piazza, and van Geemen in terms of syzygy conditions, which allowed me to then further generalize it.

I am also very grateful to Riccardo Salvati Manni for bringing to my attention his work \cite{SM} and the literature on polynomials $P(Nm)$ corresponding to even cosets, and for the encouragement in exploring whether these can be used to obtained an ansatz. I am very thankful to Riccardo Salvati Manni for pointing out and giving a rigorous proof in \cite{SMnew} that $P_{i,s}^{(g)}$ are only known to be modular forms for $2^is=2^k$ for $k\ge 4$ (proposition \ref{Pismodular}), and that $P_{i,s}[\Delta]$ are modular with respect to conjugates of $\Gamma_g(1,2)$ rather than $\Gamma_g(1,2)$ itself, and for comments on the manuscript.


\begin{thebibliography}{}
\bibitem{BM} Beilinson, A., Manin, F.: {\it The Mumford form and the Polyakov measure in string theory.} Commun. Math. Phys. {\bf 107} (1986) 359--376.
\bibitem{CDP} Cacciatori, S.L., Dalla Piazza, F.: {\it Two loop superstring amplitudes and $S_6$ representations.} Lett. Math. Phys. (2008); arXiv:0707.0646.
\bibitem{CDPvG} Cacciatori, S.L., Dalla Piazza, F., van Geemen, B.: {\it Modular Forms and Three Loop Superstring Amplitudes.} arXiv:0801.2543.
\bibitem{vgvdg}van Geemen, B., van der Geer, G.: {\it Kummer
    varieties and the moduli spaces of abelian varieties}, Amer. J.
    of Math. {\bf 108} (1986) 615--642.
\bibitem{GS}Green, M.B., Schwarz, J.H.: {\it Supersymmetrical string theories.} Phys. Lett. B {\bf 109} (1982) 444--448.
\bibitem{GHMR}Gross, D.J., Harvey, J.A., Martinec, E.J., Rohm, R. {\it Heterotic String Theory (II). The interacting heterotic string.} Nucl. Phys. B {\bf 267} (1986) 75.
\bibitem{DHP86} D'Hoker, E., Phong, D.H.: {\it Multiloop amplitudes for the bosonic Polyakov string}, Nucl. Phys. B {\bf 269} (1986) 205--234.
\bibitem{DHP1} D'Hoker, E., Phong, D.H.: {\it Two-Loop
    Superstrings I, Main Formulas.} Phys. Lett. B {\bf 529} (2002)
    241--255; hep-th/0110247.
\bibitem{DHPa} D'Hoker, E., Phong, D.H.: {\it Two-Loop Superstrings II, The chiral Measure on Moduli Space.} Nucl. Phys. B {\bf 636} (2002) 3--60; hep-th/0110283.
\bibitem{DHPb} D'Hoker, E., Phong, D.H.: {\it Two-Loop Superstrings III, Slice Independence and Absence of Ambiguities.} Nucl. Phys. B {\bf 636} (2002) 61--79; hep-th/0111016.
\bibitem{DHPc} D'Hoker, E., Phong, D.H.: {\it Two-Loop Superstrings IV, The Cosmological Constant and Modular Forms.} Nucl. Phys. B {\bf 639} (2002) 129--181; hep-th/0111040.
\bibitem{DHP2} D'Hoker, E., Phong, D.H.: {\it Asyzygies, modular forms, and the superstring measure I.} Nucl. Phys. B {\bf 710}, 58 (2005); hep-th/0411159.
\bibitem{DHP3} D'Hoker, E., Phong, D.H.: {\it Asyzygies, modular forms, and the superstring measure. II.} Nucl. Phys. B {\bf 710}, 83 (2005); hep-th/0411182.
\bibitem{DHPd} D'Hoker, E., Phong, D.H.: {\it Two-Loop Superstrings V, Gauge Slice Independence of the N-Point Function.} Nucl. Phys. B {\bf 715} (2005); hep-th/0501196.
\bibitem{DHPe} D'Hoker, E., Phong, D.H.: {\it Two-Loop Superstrings VI, Non-Renormalization Theorems and the 4-Point Function.} Nucl. Phys. B {\bf 715} (2005); hep-th/0501197.
\bibitem{I}Igusa, J.-I.: Theta functions. Die Grundlehren
    der mathematischen Wissenschaften, Band 194. Springer-Verlag, New    York-Heidelberg, 1972.
\bibitem{K}Krazer, A.: Lehrbuch der Thetafunktionen, B. G. Teubner, Leipzig, 1903.
\bibitem{Manin} Manin, Y.,{\it The partition function of the Polyakov string can be expressed in terms of theta functions}, Phys. Lett. B {\bf 172} (1986) 184--185.
\bibitem{MV}Matone, M., and Volpato, R. {\it Higher genus superstring amplitudes from the geometry of moduli space.} Nucl. Phys. B {\bf 732} (2006) 321--340; hep-th/0506231.
\bibitem{OPY} Oura, M., Poor, C., Yuen, D.S., {\it Toward the Siegel ring in genus four,} Int. J. Number Th., to appear.
\bibitem{OSM}Oura, M., Salvati Manni, R.: {\it On the image of code polynomials under theta map}, preprint.
\bibitem{PY}Poor, C., Yuen, D.S., {\it Linear dependence among Siegel modular forms.}  Math. Ann.  {\bf 318}  (2000) 205--234.
\bibitem{SM} Salvati Manni, R.: {\it On the dimension of the vector space $\C[\theta_m]_4$}, Nagoya Math. J. {\bf 98} (1985) 99--107.
\bibitem{SMlevel}Salvati Manni, R.: {\it Modular varieties with
    level 2 theta structure}, Amer. J. Math. {\bf 116} (1994)
    1489--1511.
\bibitem{SMnew}Salvati Manni, R.: {\it Remarks on Superstring  amplitudes in higher genus}, Nucl. Phys. B, to appear, arXiv:0804.0512.
\bibitem{VV}Verlinde, E., Verlinde, H. {\it Chiral Bosonization, determinants and the string partition function.} Nucl. Phys. B {\bf 288} (1987) 357--396.
\end{thebibliography}
\end{document}